\documentclass[a4paper,11pt]{article}
\usepackage[left=2cm,right=2cm,top=2cm,bottom=2cm]{geometry}
\usepackage[usenames,dvipsnames,svgnames,table]{xcolor}
\usepackage{amsmath,lscape,amsfonts,amssymb,amsthm,verbatim,lscape}
\usepackage[figuresright]{rotating}
\usepackage{natbib}
\usepackage{url}
\usepackage{hyperref}
\hypersetup{colorlinks,
citecolor=black,
filecolor=black,
linkcolor=black,
urlcolor=black,
pdftex}

\usepackage{booktabs}

\usepackage{colortbl}
\definecolor{gray}{rgb}{0.8,0.8,0.8}

\usepackage{times}
\usepackage{blindtext}
\usepackage[font={footnotesize,it}]{caption}

\usepackage{sectsty}  
\subsectionfont{\normalsize\bf}
\sectionfont{\large\bf}

\def\Rtil{{\widetilde{\R}}}
\def\atil{{\widetilde\a}}

\newcommand{\bea}{\begin{eqnarray}}
\newcommand{\eea}{\end{eqnarray}}
\def\bbf{\boldsymbol{\beta}}

\def\gbf{\boldsymbol{\gamma}}

\def\th{\theta}
\def\thbf{\boldsymbol{\theta}}

\def\Debf{\boldsymbol{\Delta}}
\def\E{{\rm E}\,}

\def\a{\mathbf{a}}
\def\g{\mathbf{g}}

\def\s{\mathbf{s}}

\def\V{\mathbf{V}}
\def\bfD{\mathbf{H}}

\def\R{\mathbf{R}}

\def\M{\mathbf{J}}

\def\W{\mathbf{W}}
\def\x{\mathbf{x}}
\def\X{\mathbf{X}}
\def\Y{\mathbf{Y}}
\def\y{\mathbf{y}}
\def\Rtil{{\widetilde{\R}}}

\def\b{\beta}

\def\O{{\boldsymbol\Omega}}

\def\D{{\Delta}}
\def\p{\partial}

\usepackage{titlesec}
\titlespacing\subsubsection{0pt}{2pt plus 4pt minus 2pt}{2pt plus 2pt minus 2pt}

\begin{document}

\title{Weighted scores method for longitudinal ordinal data}

\author{
Aristidis K. Nikoloulopoulos\footnote{{\small\texttt{A.Nikoloulopoulos@uea.ac.uk}}, School of Computing Sciences, University of East Anglia,
Norwich NR4 7TJ, UK}
}
\date{}

\maketitle

\begin{abstract}
\baselineskip=16pt
\noindent
Extending generalized estimating equations (GEE) to ordinal response data
requires 
a conversion of the ordinal response to a vector of binary category indicators. That leads to a rather complicated association structure, and the introduction of large matrices when the number of categories and dimension of the cluster are large. 
To 
allow a richer specification of working correlation assumptions, we adopt the
weighted scores method  which is essentially an extension  of the
GEE approach, since it can also be applied to families that are not in
the GLM class. 
The weighted scores method  stems from the lack of a theoretically sound methodology for analyzing multivariate discrete data based only on moments up to second order and  it  is robust to dependence and nearly as efficient as maximum likelihood.  There is 
no need to convert the ordinal response to binary indicators, thus the weight
matrices have smaller dimensions and it is not necessary to guess the
correlations of indicator variables for different categories. We focus on  important issues that would interest the data analyst, such as
choice of the structure of the correlation matrix and  of explanatory variables,
comparison of results obtained from our methods versus GEE,  and  insights provided by our method that would be missed with the GEE method. Our modelling framework is implemented in the package {\tt weightedScores} within the open source statistical environment {\tt R}. 
\\\\
\noindent {\it Keywords:} AIC/BIC; Composite likelihood; Correlation structure selection;   Generalized estimating equations; Ordinal regression; Variable selection.
\end{abstract}

\baselineskip=16pt

\section{\label{intro}Introduction}
The method of generalized estimating equations   
\citep[GEE hereafter]{liang&zeger86,zeger&liang86}, which is  popular in biostatistics,
analyzes correlated data by assuming a generalized linear model (GLM)
for the outcome variable, and a structured correlation matrix to describe the pattern of association
among the repeated measurements on each subject or cluster. The associations  are treated as nuisance parameters; interest focuses  on the statistical inference for the regression parameters and the method is based only on moments up to second order.

Extending GEE to ordinal response data, say with $K$ categories,
requires an alteration of the general theory because the first and second moments are not defined for ordinal observations. This modification is based on
a conversion of the ordinal response to a vector of $K-1$ binary indicators of categories $1,\ldots,K-1$ \citep{lipsitz-etal-94,heagerty&zeger1996,parsons-etal-2006,touloumis-agresti-kateri-2013}. 
There are various options for choosing the binary variables  and also various parts of associations which will eventually describe all of the possible outcomes for the original ordinal responses. The first part is the association between the binary variables at one time point. The second is the association of the same coded binary variables across time, and the third and final part is the association of two differently coded binary variables across time \citep{Nooraee-etal-2014}. This leads to a rather complicated association structure and the introduction of large matrices when $K$ and $d$ are large, where $d$ is the dimension of a ``cluster" or
``panel".

{\cite{Parsons-2013-SIM} discussed many  realistic examples, where  repeated ordinal  data with a large number of categories  are involved,  e.g., clinical scoring systems such as the Oxford Hip Score \citep{Dawson1996},  and used the 
Warwick Arthroplasty Trial \citep{Achten-etal-2010}  data with $K=49$ categories.}
{\cite{Parsons-2013-SIM}} described these as long ordinal scores or composite ordinal scores, that result from complex surveys. When $K$ or $d$ is large the working correlation matrix, a square matrix of dimension $d(K-1)$, is very large, as it needs to account for correlations between the $K$-1 new binary scores at each time-point and between time-points. Matrix operations (e.g. inversion) required for parameter estimation can become very slow or even infeasible for large $K$ or $d$. For long ordinal scores, with a large number of categories, this presents a problem, as many cut-point parameters would need to be estimated with presumably poor precision and likely convergence problems that are often particularly associated with models for repeated ordinals scores {\citep{Lipsitz-etal-1994-Biometrics,parsons-etal-2006,Parsons-2013-SIM,touloumis-agresti-kateri-2013,Nooraee-etal-2014}}.

{\cite{Nooraee-etal-2014} made much of convergence issues with available packages for model fitting in this setting. This study used only the default options in existing softwares. This does not necessarily imply that convergence issues will still be present if the maximum number of iterations is increased. Our own take  on this issue  is rather different from those expressed in this latter paper. Of course quite what we mean by lack of convergence is also often not clearly defined -- simply changing a starting value(s) for the association parameter(s), which is (are) often regarded as problematic, can often simply solve the issue.  Hence we don't   associate lack of convergence for GEE with large matrices. 
However, large matrices are a problem for GEE methods, in as much as they cause model fitting algorithms to run slowly.  }

{Furthermore, \cite{Nooraee-etal-2014} provided a comprehensive comparison of existing GEE approaches for ordinal data and  revealed that different methods can lead to different estimates and identified the local odds ratio GEE approach in \cite{touloumis-agresti-kateri-2013} as the dominant among the existing  GEE methods. 
Note in passing that} each parametrization measures the correlation in a
different way, e.g., using correlation coefficients, local odds ratios and global odds
ratios (see \cite{touloumis-agresti-kateri-2013}, and the references therein). Therefore different
parametrizations would indeed lead to slightly different estimates in finite samples for the
existing GEE methods but according to the GEE theory developed by \cite{liang&zeger86} all of them should produce consistent estimators of the marginal regression vector provided that the estimator of the association parameter vector is
$\sqrt{N}$-consistent given the true regression parameter vector.

 \cite{nikoloulopoulos&joe&chaganty10} developed the weighted scores method for regression models with dependent data. 
 Similar to GEE which constructs unbiased equations weighting the residuals, the weighted scores method constructs estimating equations weighting the univariate score functions, placing the weights strategically. The novelty is the use of a discretized multivariate normal  distribution as a working model to specify the weights of estimating equations based on the univariate marginal distributions of the response. Thus it can be regarded as a generalization of GEE in the sense that our method is applicable to a wider family of regression models that are not necessarily in the class of generalized linear models.  
For concreteness,
the theory was illustrated for discrete negative binomial margins, that are not in the GLM
family.

The weighted scores method  stems from the lack of a theoretically sound methodology for analyzing multivariate discrete data based only on moments up to second order, and in part from recent criticism  of the GEE method \citep{Lindsey&Lambert98,chaganty&joe06,touloumis-agresti-kateri-2013}.  The
weighted scores method is  based on a plausible discretized multivariate normal (MVN) model, and the
nuisance parameters  are interpretable as latent correlation
parameters.  This avoids problems of interpretation in GEE for a working
correlation matrix that in general cannot be a correlation matrix of
the multivariate discrete  data as the univariate means change \citep{chaganty&joe06,sabo&chaganty10}. 
Further, with the GEE methodology applied to correlated discrete responses (binary, multinomial, Poisson, etc.),
 the parametric space of the pairwise association parameters is likely to be restricted by the marginal model specification \citep{bergsma&rudas2002,chaganty&joe06}. 
The ``working correlation" used by GEE lacks a proper definition relating the working correlation matrix to the probability distribution of the response vector, leading to misapplied asymptotic theory \citep{Crowder95}. With no formal definition, the working correlation, when it is not the true correlation, has no mathematical relationship to the covariance of the response vector, and in the absence of a proper underlying probability distribution assertions of consistency are invalid (law of large numbers assumes that there is an underlying probability distribution); see  \cite{lee&nelder09}. 
The weighted scores method is based on weighting the univariate score functions using a working model that is actually a proper multivariate model. Thus does not only  generalize, but also overcomes the theoretical flaws associated with GEE applied to correlated discrete responses. 
{Note that the local odds ratios GEE method in  \cite{touloumis-agresti-kateri-2013} is estimating the nuisance parameter vector by maximizing an objective function \citep{Crowder95} and hence also avoids the aforementioned pitfalls in the GEE method. Their  model with log odds implies an underlying bivariate Plackett \citep{plackett65} copula  \citep[Supplementary Materials]{touloumis-agresti-kateri-2013}. The multivariate Plackett has not been proved to be a proper multivariate  model, but   \cite{chaganty&joe06} showed  that the range of bivariate 
log odds ratios is not as constrained as bivariate correlations for
multivariate binary. }

The weighted scores method 
has the merit of robustness to misspecification of the dependence structure like in GEE, but with the additional advantage, with respect to GEE, that dependence is expressed in terms of a ``real" multivariate model. 
 \cite{nikoloulopoulos&joe&chaganty10} demonstrated with theoretical  and simulation  studies that   the weighted scores method is highly efficient when compared with the ``gold standard'' maximum likelihood  methods and robust to joint distribution assumptions. The estimating equations based on a ``working" MVN copula-based model are robust if the univariate model is correct, while on the other hand ML estimates could be biased if the univariate model is correct but dependence is modelled incorrectly.

Model selection is  an important  issue in longitudinal data analysis, since 
when conducting a GEE analysis, it is essential  to carefully model the correlation parameters, in order to avoid a substantive loss in efficiency in the estimation of the regression parameters \citep{Albert&McShane1995,Crowder95,wang&carey2003,Shults-etal2009}.
\cite{Nikoloulopoulos2015d} has  proposed the CL1 information  criteria as an intermediate step for correlation structure  and variable selection in the weighted scores method. The proposed criteria have the similar attractive property with QIC \citep{pan2001} of allowing covariate selection and working correlation structure selection using the same model selection criteria. It has been demonstrated that outperform  QIC and several other existing  approaches in the GEE literature for model selection. The main reason is that they are being likelihood-based \citep{Varin-etal2011}.

In this paper we modify the general theory in 
 \cite{nikoloulopoulos&joe&chaganty10} and 
  \cite{Nikoloulopoulos2015d} for  longitudinal ordinal response data. 
To sum up the advantages of the proposed method over existing approaches in GEE for ordinal regression are (a) the avoidance of large matrices when $K$ or $d$ are large (since there is no need to create a set of binary variables), 
 (b)  a proper definition relating the working disretized MVN model to the probability distribution of the response vector, (c) a latent  correlation matrix for the ordinal outcomes induced by the MVN latent variables, and (d) analogues of the AIC and the BIC for  variable and correlation  structure selection, namely the CL1AIC and CL1BIC, can be derived. 
Hence the weighted scores method for repeated ordinal responses  on the one hand  can allow for a richer specification of the correlation   assumptions  and on the other hand  a  correct specification in both  the correlation and mean function (covariate)  modelling.

The remainder of the paper proceeds as follows. Section \ref{sec-wtsc} provides the theory of the weighted scores method for ordinal  regression with dependent data. Section  \ref{cl1-sec} derives the CL1  information criteria in the context of longitudinal data analysis with an ordinal  margin.
Section \ref{sec-sim} describes the simulation studies we perform to gauge the efficiency and robustness  of the weighted scores and GEE method, and to assess the performance of the CL1 information criteria for longitudinal ordinal  data.
We discuss an application example in Section \ref{sec-app}
and conclude with some discussion in Section \ref{sec-disc}, followed by a brief section with the software details.

\section{\label{sec-wtsc}The weighted scores estimating equations}

The main idea of weighted scores method is to write out
the score equations for independent data within clusters
or panels, and then generalize to estimating equations by
inserting weight matrices between the matrix of covariates and the  vector
of scores for regression and non-regression parameters.
The general theory \citep{nikoloulopoulos&joe&chaganty10} is modified here for ordinal regression models.

For ease of exposition, let $d$ be the dimension of a ``cluster" or
``panel" and $n$ the number of clusters.  The theory can
be extended to varying cluster sizes.
Let $p$ be the number of
covariates, that is, the dimension of a covariate vector $\x$.
Let $Z\sim \mathcal{F}$ be a latent variable,  such that $Y=y$ if
$\alpha_{y-1}+\x^T\bbf\leq Z\leq  \alpha_{y}+\x^T\bbf,\,y=1,\ldots,K,$
where $K$ is the number of categories of $Y$
(without loss of generality, we assume $\alpha_0=-\infty$ and  $\alpha_K=\infty$), and $\bbf$ is the $p$-dimensional regression vector.
From this definition, the response $Y$ is assumed to have density
$$f_1(y;\nu,\gbf)=\mathcal{F}(\alpha_{y}+\nu)-\mathcal{F}(\alpha_{y-1}+\nu),$$
where $\nu=\x^T\bbf$ is a function of $\x$
and the $p$-dimensional regression vector $\bbf$, and $\gbf=(\alpha_1,\ldots,\alpha_{K-1})$ is the $q$-dimensional vector of the univariate cutpoints ($q=K-1$). Note that $\mathcal{F}$ normal leads to the probit model and $\mathcal{F}$ logistic
leads to the cumulative logit model for ordinal response.

Suppose that the data are $(y_{ij},\x_{ij})$, $j=1,\ldots,d$, $i=1,\ldots,n$,
where $i$ is an index for individuals or clusters, $j$ is an index for
the repeated measurements or within cluster measurements.
The univariate marginal model for
$Y_{ij} $ is $f_1(y_{ij}; \nu_{ij},\gbf)$ where  $\nu_{ij}=\x_{ij}^\top\bbf$ and  $\gbf$ of dimension $q$ be the vector of univariate cutpoints. {Here we  consider univariate  parameters that are common to different margins, i.e., common regression parameters $\bbf$ and cut-points $\gbf$ for different univariate margins. The theory can
be extended for the case   the univariate parameters are not common to the different margins.} If for each $i$, $Y_{i1},\ldots,Y_{id}$ are independent, then the
log-likelihood is
$$L_1= \sum_{i=1}^n\sum_{j=1}^d\, \log f_1(y_{ij};\nu_{ij},\gbf)=\sum_{i=1}^n\sum_{j=1}^d\,
 \ell_1(\nu_{ij},\gbf,\, y_{ij}),$$
where $\ell_1(\cdot)=\log \, f_1(\cdot)$. The score equations for $\bbf$ and $\gbf$ are
\bea\label{gs3}
\begin{pmatrix}\frac{\p L_1}{\p \bbf}\\\frac{\p L_1}{\p \gbf}\end{pmatrix}=
\sum_{i=1}^n\sum_{j=1}^d\, \begin{pmatrix}
\x_{ij} & {\bf 0} \cr
{\bf 0} & {\bf I}_q \end{pmatrix} \begin{pmatrix}
\frac{\p \ell_1(\nu_{ij},\gbf,\, y_{ij})}{\p\nu_{ij}} \cr
\frac{\p\ell_1(\nu_{ij},\gbf,\, y_{ij})}
 {\p\gbf} \end{pmatrix}=
\sum_{i=1}^n\sum_{j=1}^d\,
\begin{pmatrix}
\x_{ij}\mathbf{1}_q \cr
 {\bf I}_q \end{pmatrix}
 \frac{\p \ell_{1ij}(\gbf_{ij}, y_{ij})}{\p\gbf_{ij}}
={\bf 0},
 \eea
where $\gbf_{ij}=(\alpha_1+\nu_{ij},\ldots,\alpha_{K-1}+\nu_{ij})=(\gamma_{ij1},\ldots,\gamma_{ij,K-1})$, $\ell_{1ij}(\cdot)=\log f_{1ij}(\cdot), f_{1ij}(\gbf_{ij}, y)=\mathcal{F}(\gamma_{ijy})-\mathcal{F}(\gamma_{ij,y-1})$,  and ${\bf I}_q$ is an identity matrix of
dimension $q$.
Let $\X_{ij}^T=\begin{pmatrix}
\x_{ij}\mathbf{1}_q \cr
 {\bf I}_q \end{pmatrix}$
and   $ \s_{ij}^{(1)}(\a)=\frac{\p \ell_{1ij}(\gbf_{ij}, y_{ij})}{\p\gbf_{ij}}$ where $\a^\top=(\bbf^\top,\gbf^\top)$ is the
column vector of all  $r=p+q$ univariate parameters.
The score equations (\ref{gs3}) can be written as
 \begin{equation}\label{independent score equations}
\g_1=\g_1(\a)=\frac{\p L_1}{\p \a}=  \sum_{i=1}^n\sum_{j=1}^d\X_{ij}^\top\;\s_{ij}^{(1)}(\a)=
\sum_{i=1}^n\X_i^\top\;\s_i^{(1)}(\a)={\bf 0},
\end{equation}
where
$\X_{i}^\top=(\X_{i1}^\top, \ldots, \X_{id}^\top)$ and $\s_i^{(1)\top}(\a)=
(\s_{i1}^{(1)\top}(\a),\ldots,\s_{id}^{(1)\top}(\a) )$.
The vectors  $\s_{ij}^{(1)}(\a)$ and $\s_i^{(1)}(\a)$ have dimensions $q$ and
$dq$ respectively.
The dimensions of $\X_{ij}$ and $\X_i$ are $q\times r$ and  $dq \times r$
respectively.

For estimation of $\a$,  when $Y_{i1},\ldots,Y_{id}$ are dependent and
a multivariate model is not used, an approach is to use a
``working model" for the purpose of getting the weight matrices,   
which
might be near optimal for the ``true joint distribution".
We select a  ``working model" based on the
discretized MVN distribution as this allows a wide range of dependence.
The  discretized MVN (or the multivariate normal copula with discrete margins) model has the following
cumulative distribution function (cdf):
 $$F(y_{1}, \dots, y_{d})=\Phi_d\left(\Phi^{-1}[F_1(y_{1};\nu_{1},\gbf)],\ldots,
  \Phi^{-1}[F_1(y_{d};\nu_{d},\gbf)];\R\right),$$
where $\Phi_d$ denotes the standard MVN distribution function with correlation
matrix $\R=(\rho_{jk}: 1\le j<k\le d)$, $\Phi$ is cdf of the univariate standard normal,
and $F_1(y;\nu,\gbf)=\mathcal{F}(\alpha_y+\nu)$ is the univariate
cdf for $Y$. The MVN copula inherits the
dependence structure of the MVN distribution, but lacks a closed form cdf;
this means likelihood inference might be difficult as
$d$-dimensional integration is required for the multivariate probabilities ($d>3$); see e.g.,  \cite{Nikoloulopoulos&karlis07FNM}.
However, in our case as the  weight matrices will depend on the covariances of the scores, only the bivariate
marginal probabilities of $Y_{ij}$ and $Y_{ik}$, $j\neq k$ will be needed  for estimation.

The estimating equations 
based on a ``working"
discretized MVN, take the form:
\begin{equation}\label{wtee}
  \g_1^\star=\g_1^\star(\a)=\sum_{i=1}^n\X_i^T\,\W_{i,\rm working}^{-1}\,\s_i^{(1)}(\a)={\bf 0},
\end{equation}
where $\W_{i,\rm working}^{-1}=
\D_i^{(1)}(\atil)[\O_i^{(1)}(\atil,\Rtil)]^{-1}$ is based on
the covariance matrix $\O_i^{(1)}(\atil,\Rtil)$ of $\s_i^{(1)}(\a)$ computed from the
fitted discretized MVN model with estimated parameters $\atil$ and $\Rtil$ and the symmetric
$dq\times dq$ matrix $\D_i^{(1)}=\mbox{diag}(\D_{i1}^{(1)},\ldots,\D_{id}^{(1)})$ with $\D_{ij}=-\E(\frac{\p^2 \ell_{1ij}(\gbf_{ij}, y_{ij})}{\p\gbf_{ij}\p\gbf_{ij}^\top})$. As bivariate normal cdf calculations  are needed
for the calculation of  $\O_i^{(1)}(\atil,\Rtil)$ (different ones for different clusters),
a good approximation that can be quickly computed  is important.
We used the approximation given by \cite{Johnson&Kotz72}.

The estimated  parameters   $\atil$ and $\Rtil$  of the working discretized MVN model can be easily obtained in a two-step approach, namely the CL1 method in \cite{zhao&joe05}. Estimated $\atil$ and $\Rtil$   are obtained 
by solving the CL1 univariate    and   bivariate composite score functions,  respectively. The former are the same with the independent estimating equations  (\ref{independent score equations}), while the latter are given below:
$$\g_2=\sum_{i=1}^{n}\s_i^{(2)}(\atil,\R)
 =\mathbf{0},$$
where $\s_i^{(2)}(\a,\R)=\frac{\p\sum_{j<k}\log f_2(y_{ij},y_{ik};\nu_{ij},\nu_{ik},\gbf,\rho_{jk})}{\p \R}$
 with $f_2(\cdot)$  the bivariate
marginal probability of $Y_{ij}$ and $Y_{ik}$, viz.
$$f_2(y_{ij},y_{ik};\nu_{ij},\nu_{ik},\gbf,\rho_{jk})=\int_{\Phi^{-1}[F_1(y_{ij}-1;\nu_{ij},\gbf)]}^{\Phi^{-1}[F_1(y_{ij};\nu_{ij},\gbf)]}
\int_{\Phi^{-1}[F_1(y_{ik}-1;\nu_{ik}),\gbf]}^{\Phi^{-1}[F_1(y_{ik};\nu_{ik},\gbf)]}  \phi_2(z_j,z_d;\rho_{jk}) dz_j dz_k;
$$
 $\phi_2(\cdot;\rho)$ denotes the standard bivariate normal density  with correlation
 $\rho$.

If the $\W_{i,\rm working}$ are assumed fixed for the
second stage of solving the  weighted scores estimating equations
 (\ref{wtee}), then the asymptotic covariance matrix of the
solution $\widehat\a$ is
$$  \V_1^\star=(-\bfD_{\g_1^\star})^{-1}\M_{\g_1^\star}(-\bfD^T_{\g_1^\star})^{-1}$$
with
\bea\label{robust2}
  -\bfD_{\g_1^\star}
  =\sum_{i=1}^n\X_i^T\W_{i,\rm working}^{-1}\D_i^{(1)}\X_i,\;
  \M_{\g_1^\star} &=& \sum_{i=1}^n\X_i^T\W_{i,\rm working}^{-1}
  \O_{i,\rm true}^{(1)}(\W_{i,\rm working}^{-1})^T\X_i,\nonumber
\eea
where $\O_{i,\rm true}^{(1)}$ is the ``true covariance matrix" of $\s_i^{(1)}(\a)$. The $\O_{i,true}^{(1)}$ can be estimated by $\s_i^{(1)}(\widehat\a)\s_i^{(1)^\top}(\widehat\a)$. This estimate is similar to what is done in the ``sandwich" covariance estimator in GEE. 

\section{\label{cl1-sec}CL1 information criteria}

The CL1 method in  \cite{zhao&joe05}  is used to estimate conveniently the univariate and latent correlation parameters of the discretized MVN model in order to compute the working weight matrices and then solve the weighted score equations (\ref{wtee}). Herein, we also call the CL1  information criteria, for correlation structure and variable selection in the weighted scores estimating equations.
This section provides the form of the CL1 information criteria, proposed  by \cite{Nikoloulopoulos2015d} for longitudinal binary and count data, in the context of longitudinal  ordinal data.

The CL1 versions of AIC and  BIC criteria are defined as:
$$
\mbox{CL1AIC} =  -2L_2 + 2\mbox{tr}\Bigl(\M_\g\bfD_\g^{-1}\Bigr);
$$
$$
\mbox{CL1BIC} =-2L_2+\log(n)\mbox{tr}\Bigl(\M_\g\bfD_\g^{-1}\Bigr),
$$
where $L_2=\sum_{i=1}^{n}\sum_{j<k}\log f_2(y_{ij},y_{ik};\nu_{ij},\nu_{ik},\gbf,\rho_{jk})$
is the bivariate CL1 log-likelihood, $\M_\g$ is the covariance or variability matrix, and $\bfD_\g$ is the sensitivity or Hessian matrix of the CL1 estimating equations  $\g=(\g_1,\g_2)^\top$. 

The covariance matrix $\M_\g$ of the composite score functions $\g$ is given as below
$$\M_\g=\mbox{Cov}(\g)=
\begin{pmatrix}\mbox{Cov}(\g_1) & \mbox{Cov}(\g_1,\g_2)\\
\mbox{Cov}(\g_2,\g_1) & \mbox{Cov}(\g_2)
\end{pmatrix}=
\frac{1}{n}\sum_i\begin{pmatrix}\X_i^\top\O_i^{(1)}\X_i& \X_i^\top\O_i^{(1,2)}\\
\O_i^{(2,1)}\X_i& \O_i^{(2)}\end{pmatrix},
$$
where
$$\begin{pmatrix}\O_i^{(1)}& \O_i^{(1,2)}\\
\O_i^{(2,1)}& \O_i^{(2)}\end{pmatrix}=
\begin{pmatrix}\mbox{Cov}\Bigl(\s_i^{(1)}(\a)\Bigr) & \mbox{Cov}\Bigl(\s_i^{(1)}(\a),\s_i^{(2)}(\a,\R)\Bigr)\\
\mbox{Cov}\Bigl(\s_i^{(2)}(\a,\R),\s_i^{(1)}(\a)\Bigr) & \mbox{Cov}\Bigl(\s_i^{(2)}(\a,\R)\Bigr)
\end{pmatrix}.$$

To define the  Hessian matrix of the CL1 estimating equations, first set $\thbf=(\a,\R)^\top$,  then
$$
-\bfD_\g=E\Bigl(\frac{\p \g}{\p\thbf}\Bigr)=
\begin{pmatrix}
E\Bigl(\frac{\p \g_1}{\p\a}\Bigr)&E\Bigl(\frac{\p \g_1}{\p\R}\Bigr)\\
E\Bigl(\frac{\p \g_2}{\p\a}\Bigr)&E\Bigl(\frac{\p \g_2}{\p\R}\Bigr)
\end{pmatrix}=
\begin{pmatrix}
-\bfD_{\g_1}&\mathbf{0}\\
-\bfD_{\g_{2,1}}&-\bfD_{\g_2}
\end{pmatrix},
$$
where $-\bfD_{\g_1}=\frac{1}{n}\sum_i^n\X_i^\top\Debf_i^{(1)}\X_i$,
$-\bfD_{\g_{2,1}}=\frac{1}{n}\sum_i^n\Debf_i^{(2,1)}\X_i$,
and
$-\bfD_{\g_2}=\frac{1}{n}\sum_i^n\Debf_i^{(2,2)}$. The forms of  $\Debf_i^{(1)},\Debf_i^{(2,1)},\Debf_i^{(2,2)}$ are given in an Appendix.

Summing up, to evaluate the CL1 information criteria  involves the computation  of matrices described above; their dimensions  are given in Table \ref{dims}. The computations involve the trivariate and four-variate margins along with their derivatives;  technical details are shown in the Appendix.

\setlength{\tabcolsep}{6pt}
\begin{table}[!h]
\centering
\caption{\label{dims}The dimensions of various matrices involved in the calculation of CL1 information criteria. Note that $t=p + \binom{d}{2}+q$ and $r=p+q$.}
\begin{tabular}{cccccccccc}
\toprule Matrix &$\M_\g$ &  $\M_\g^{(1)}$ &  $\M_\g^{(1,2)}$ &  $\M_\g^{(2,1)}$ & $\M_\g^{(2)}$ & $\O_i^{(1)}$ & $\O_i^{(1,2)}$ &$\O_i^{(2,1)}$ &$\O_i^{(2)}$\\
Dimensions & $t\times t $ &$r\times r$ &$r\times \binom{d}{2}$ & $ \binom{d}{2} \times r$ &$\binom{d}{2}\times\binom{d}{2}$&$dq\times dq$ &$dq\times\binom{d}{2}$ &$\binom{d}{2}\times dq$ &$\binom{d}{2}\times\binom{d}{2}$

\\\hline
Matrix & $\bfD_\g$ & $\bfD^{(1)}$ & $\bfD^{(1,2)}$ & $\bfD^{(2,1)}$ & $\bfD^{(2)}$  & $\Debf_i^{(1)}$ & $\Debf_i^{(1,2)}$ &
$\Debf_i^{(2,1)}$ & $\Debf_i^{(2)}$\\
Dimensions & $t\times t$ & $r\times r$ &  $r\times \binom{d}{2}$ & $ \binom{d}{2} \times r$& $\binom{d}{2}\times\binom{d}{2}$
&$dq\times dq$ &$dq\times\binom{d}{2}$ &$\binom{d}{2}\times dq$ &$\binom{d}{2}\times\binom{d}{2}$  \\

\bottomrule
\end{tabular}
\end{table}

\section{\label{sec-sim}Simulations}
In order to study the robustness and efficiency of the weighted scores method for longitudinal ordinal responses,  we will use various multivariate copula models as true models. 
We will compare  the weighted scores method with  the `gold standard' maximum likelihood and also include in the comparison the local odds ratio GEE  approach  in 
 \cite{touloumis-agresti-kateri-2013} as the current state of the art of the various GEE approaches for ordinal regression \citep{Nooraee-etal-2014}.  This  GEE approach avoids the theoretical problems (see Section \ref{intro}) by using a local odds ratios parametrization to describe the association pattern within subjects. 
We also assess the performance of the CL1 information criteria proposed by  
\cite{Nikoloulopoulos2015d}  for longitudinal ordinal response data. Before that, the first subsection provides some background on copula models that might be suitable for clustered and longitudinal ordinal data.

\subsection{Relevant background for copula models}
A copula is a multivariate cdf with uniform $U(0,1)$ margins \citep{joe97,joe2014,nelsen06}. 
If $\mathcal{G}$ is a $d$-variate cdf with univariate margins $\mathcal{G}_1,\ldots,\mathcal{G}_d$,
then Sklar's (1959) \nocite{sklar1959} theorem implies that there is a copula $C$ such that
  $$\mathcal{G}(y_1,\ldots,y_d)= C\Bigl(\mathcal{G}_1(y_1),\ldots,\mathcal{G}_d(y_d)\Bigr).$$
{Copulas enable you to break
the  model building process   into two separate
steps:}

{
\begin{enumerate}
\itemsep=0pt
\item Choice of arbitrary marginal distributions $\mathcal{G}_1(y_1),\ldots,\mathcal{G}_d(y_d)$;
\item Choice of an arbitrary copula function $C$ (dependence structure).
\end{enumerate}}
\vspace{-2ex}

{\noindent If one assumes a different copula, then a different  multivariate distribution is constructed.}
The copula is unique if $\mathcal{G}_1,\ldots,\mathcal{G}_d$ are continuous, but not
if some of the $\mathcal{G}_j$ have discrete components.
If $\mathcal{G}$ is continuous and $(Y_1,\ldots,Y_d)\sim \mathcal{G}$, then the unique copula
is the distribution of $(U_1,\ldots,U_d)=\left(\mathcal{G}_1(Y_1),\ldots,\mathcal{G}_d(Y_d)\right)$ leading to
  $$C(u_1,\ldots,u_d)=\mathcal{G}\Bigl(\mathcal{G}_1^{-1}(u_1),\ldots,\mathcal{G}_d^{-1}(u_d)\Bigr),
  \hspace{2ex} 0\le u_j\le 1, j=1,\ldots,d,$$
where $\mathcal{G}_j^{-1}$ are inverse cdfs. In particular,
if $\mathcal{T}_d(\cdot;\nu,\R)$
is the MVT cdf with correlation  matrix $\R=(\rho_{jk}: 1\le j<k\le d)$ and $\nu$  degrees of freedom, and $\mathcal{T}(\cdot;\nu)$ is the univariate Student t cdf with $\nu$  degrees of freedom,
then the MVT copula is

$$C(u_1,\ldots,u_d)=\mathcal{T}_d\Bigl(\mathcal{T}^{-1}(u_1;\nu),\ldots,\mathcal{T}^{-1}(u_d;\nu);\nu,\R\Bigr).
$$

For ordinal (discrete) random vectors, multivariate probabilities
of the form $\pi_d(\y)=\Pr(Y_1=y_1,$ $\ldots,
Y_d=y_d)$  involve $2^d$ finite
differences of the joint cdf.
Therefore  likelihood inference for discrete data is straightforward 
for copulas with a computationally feasible form of the cdf.
Archimedean \citep{joe97} and mixtures
 of max-id \citep{joe&hu96} parametric family of copulas have closed form cdfs but
have less range of dependence compared with the MVN or MVT copulas.
However they provide  enough structure to study
the efficiency of the weighted scores method using the
discretized MVN as a ``working model". For example,
the Archimedean copula is suitable for positive dependent
clustered data with exchangeable dependence, while the mixture
of max-id copula is suitable for more general positive
dependence, including dependence that is decreasing with
lag as in longitudinal data. {More importantly, these copulas  have different  dependence properties than the ``working" MVN copula. For example they provide reflection  asymmetric tail dependence, while the MVN copula provides tail independence \citep{joe97,joe2014}. Hence they are suitable to study the robustness to dependence of the weighted scores method. These parametric families of copulas are briefly defined below:}

\begin{itemize}
\itemsep=0pt
\item Multivariate Archimedean copulas
have the form
$$  C(u_1, \dots, u_d\, ;\, \theta)=\phi\left(\sum^{d}_{j=1}
  \phi^{-1}(u_j\, ;\,\theta)\, ;\, \theta\right),
$$
where $\phi(u\, ;\,\theta)$ is the Laplace transform of a univariate
family of distributions of positive random variables indexed by the
parameter $\theta$, such that
$\phi(\cdot; \theta)$ and its inverse has closed form \citep{joe97}.

\item Mixture of max-id copulas \citep{joe&hu96} have  the form
\begin{multline*}C(u_1, \dots, u_d; \theta, \theta_{jk}: 1\le j<k\le d) =\\
  \phi\left(-\sum_{j<k}\log C_{jk}^{(m)}
  \bigl(e^{-p_j\phi^{-1}(u_j;\theta)},e^{-p_k\phi^{-1}(u_k;\theta)};
   \theta_{jk}\bigr)\, ; \,\theta\right); \, p_j=(d-1)^{-1},\,j=1,\ldots,d.
\end{multline*}
Since the mixing operation introduces dependence
this new copula has a dependence structure that comes from the
form of the max-id copula $C^{(m)}_{jk}(\cdot \, ;\, \theta_{jk})$ and the form of Laplace transform
$\phi(\cdot\, ;\,\theta)$. Another interesting interpretation is that the
Laplace transform $\phi$ introduces the smallest dependence between
random variables (exchangeable dependence), while the copulas $C^{(m)}_{jk}$
add some pairwise dependence. 
\end{itemize}

To this end, we consider a multivariate ordinal regression setting in which the $d \geq 2$ dependent ordinal variables $Y_{1}, \ldots, Y_{d}$ are observed together with a vector $\mathbf{x} \in \mathbb{R}^p$ of explanatory variables. If $C(\cdot)$ is any   parametric family of copulas and $F_{1}(y_j,\nu_j,\gbf)$ 
is the   parametric model for the $j$th univariate ordinal  variable  then   $$C\Bigl(F_{1}(y_{1},\nu_1,\gbf),\ldots,F_{1}(y_{d},\nu_d,\gbf)\Bigr)$$ is a multivariate parametric model with univariate margins $F_{1}(y_{1},\nu_1,\gbf),\ldots,F_{1}(y_{d},\nu_d,\gbf)$. For copula models, the response vector $\Y=(Y_1,\ldots,Y_d)$ can be  discrete \citep{Nikoloulopoulos2013a,nikoloulopoulos&joe12}
. {The only assumption we make is  that the margins of the joint distribution $\mathcal{G}$  are identical, that is, $\mathcal{G}_1=\ldots=\mathcal{G}_d=F_{1}=\mathcal{F}$. $\mathcal{F}$ normal leads to the probit model and $\mathcal{F}$ logistic
leads to the cumulative logit model for ordinal response.
The theory of the weighted scores method is not robust to margin misspecification, but it   extends to any univariate regression method. 
For instance, \cite{nikoloulopoulos&joe&chaganty10} used  discrete negative binomial margins (that are not in the GLM family) and \cite{Nikoloulopoulos2015d} used GLM (Bernoulli and Poisson) margins.}

\subsection{\label{sim-est}Small sample efficiency of the   weighted scores}
We randomly generate $B=10^4$ samples of size  {$n = 50, 100, 300$} from the  the above copula models  with exchangeable and unstructured dependence. Note  that AR(1)-like dependence is not used here since the local odds ratio GEE method in 
\cite{touloumis-agresti-kateri-2013} does not include this structure. For exchangeable dependence structure,
the Gumbel copula in the Archimedean class
 with Laplace transform
$\phi_G(t;\th)=\exp(-t^{1/\th})$ was used as the ``true model".
For unstructured dependence, the mixture of max-id copula
with Laplace transform
 $\phi_G(\cdot;\th)$ and the bivariate
Gumbel copula for the $C_{jk}^{(m)}(\cdot;\theta_{jk})$ was used as the ``true model". For simulation from Archimedean and mixture of max-id copulas we have used the  algorithms in 
\citet[pp. 272--274]{joe2014}.

We use   $d=3$, {$K=5$} ordinal categories (equally weighted) and ordinal probit regression. 
For the covariates and regression parameters, we use a combination of a time-stationary and a time-varying design, i.e., include 
covariates that are typically constant over time,  and correlated over time. 
More specifically, we chose $p=4,\x_{ij}=(x_{1ij},x_{2ij},x_{3ij},x_{4ij})$ with $x_{1i}\in\{0,1\}$ a group variable, $x_{2ij}$ an i.i.d. from a $d$-variate Gumbel copula with standard uniform margins and $d\times
d$ Kendall's tau association matrix with off-diagonal elements equal to 0.5,  $x_{3ij}=x_{1ij}\times x_{2ij}$, and $x_{4i}$
a uniform random
variable in the interval $[-1,1]$; $\beta_1=-\beta_2=-\beta_3=-0.5, \b_4=0$. By considering the noise variable $x_{4i}$ we aim to  check the Type I error rate for inference on $H_0:\b_4=0$ (see e.g., \citealp{Larrabee-etal-2014}) based on the weighted scores, local odds ratios GEE and ML methods.

\setlength{\tabcolsep}{10pt}

\begin{sidewaystable}[htbp]
  \centering
  \caption{\label{biasmse}Small sample of sizes $n=50,100,300$ simulations ($10^4$ replications) and resultant biases, root mean square errors (RMSE), and standard deviations  (SD),
along with average theoretical SDs ($\sqrt{\bar V}$) scaled by $n$, for the
maximum likelihood of the regression parameters for the trivariate Gumbel copula (exchangeable) or the mixture of max-id copula 
 with Laplace transform
 $\phi_G$ and the bivariate
Gumbel copula for the $C_{jk}^{(m)}(\cdot;\theta_{jk})$ (unstructured)
model and ordinal probit  regression, and the weighted scores (WS) and GEE  with exchangeable or unstructured  correlation matrix.}
\begin{footnotesize}

    \begin{tabular}{ccccccccccccccccccc}
    \toprule
    &       &       &       & \multicolumn{3}{c}{$\b_1=-0.5$} &       & \multicolumn{3}{c}{$\b_2=0.5$} &       & \multicolumn{3}{c}{$\b_3=0.5$} &       & \multicolumn{3}{c}{$\b_4=0$} \\
  
          &       &       &       & WS    & GEE   & ML    &       & WS    & GEE   & ML    &       & WS    & GEE   & ML    &       & WS    & GEE   & ML \\ \cmidrule{5-7} \cmidrule{9-11}  \cmidrule{13-15}  \cmidrule{17-19}

Exch  & $n=50$ &{$n$ Bias} &{} & -1.29 & -0.73 & -0.94 &       & 1.53  & 0.70  & 0.88  &       & 1.59  & 0.83  & 1.01  &       & -0.03 & -0.04 & -0.02 \\
    $\th=3$ &       &{$n$ SD} &{} & 20.88 & 20.43 & 19.37 &       & 20.24 & 19.59 & 18.46 &       & 28.99 & 28.10 & 26.51 &       & 4.22  & 4.17  & 3.89 \\
          &       &{$n$ RMSE} &{} & 20.92 & 20.44 & 19.39 &       & 20.29 & 19.60 & 18.48 &       & 29.03 & 28.11 & 26.53 &       & 4.22  & 4.17  & 3.90 \\
          &       &{$n\sqrt{\bar V}$} &{} & 19.71 & 19.29 & 18.77 &       & 18.63 & 18.10 & 17.79 &       & 26.70 & 25.98 & 25.63 &       & 4.16  & 4.09  & 3.91 \\
          & $n=100$ &{$n$ Bias} &{} & -1.35 & -0.83 & -0.83 &       & 1.54  & 0.72  & 0.79  &       & 1.61  & 0.79  & 1.12  &       & 0.00  & 0.00  & -0.02 \\
          &       &{$n$ SD} &{} & 28.39 & 28.26 & 26.55 &       & 27.06 & 26.75 & 24.99 &       & 38.76 & 38.39 & 35.85 &       & 5.79  & 5.79  & 5.42 \\
          &       &{$n$ RMSE} &{} & 28.42 & 28.27 & 26.56 &       & 27.10 & 26.76 & 25.00 &       & 38.79 & 38.40 & 35.86 &       & 5.79  & 5.79  & 5.42 \\
          &       &{$n\sqrt{\bar V}$} &{} & 27.32 & 27.14 & 25.97 &       & 25.64 & 25.43 & 24.35 &       & 36.75 & 36.53 & 35.03 &       & 5.68  & 5.68  & 5.34 \\
          & $n=300$ &{$n$ Bias} &{} & -1.05 & -0.45 & -0.41 &       & 2.02  & 1.20  & 1.67  &       & 0.85  & -0.02 & -0.22 &       & -0.26 & -0.27 & -0.18 \\
          &       &{$n$ SD} &{} & 47.09 & 47.33 & 44.39 &       & 44.69 & 44.89 & 41.94 &       & 63.70 & 64.11 & 59.82 &       & 9.64  & 9.72  & 9.08 \\
          &       &{$n$ RMSE} &{} & 47.10 & 47.33 & 44.39 &       & 44.73 & 44.91 & 41.97 &       & 63.71 & 64.11 & 59.82 &       & 9.64  & 9.72  & 9.08 \\
          &       &{$n\sqrt{\bar V}$} &{} & 46.78 & 46.85 & 44.40 &       & 43.59 & 43.74 & 41.29 &       & 62.65 & 62.95 & 59.46 &       & 9.64  & 9.73  & 9.09 \\\hline
    Exch  & $n=50$ &{$n$ Bias} &{} & -1.59 & -0.87 & -0.88 &       & 1.66  & 0.65  & 0.82  &       & 2.11  & 1.15  & 1.21  &       & -0.05 & -0.04 & -0.05 \\
    $\th=5$ &       &{$n$ SD} &{} & 19.54 & 18.80 & 18.03 &       & 16.14 & 15.14 & 14.48 &       & 23.02 & 21.56 & 20.65 &       & 3.10  & 2.99  & 2.83 \\
          &       &{$n$ RMSE} &{} & 19.60 & 18.82 & 18.05 &       & 16.23 & 15.16 & 14.50 &       & 23.11 & 21.59 & 20.69 &       & 3.10  & 2.99  & 2.83 \\
          &       &{$n\sqrt{\bar V}$} &{} & 18.50 & 17.70 & 17.54 &       & 14.77 & 13.68 & 13.95 &       & 21.22 & 19.59 & 20.18 &       & 3.03  & 2.89  & 2.82 \\
          & $n=100$ &{$n$ Bias} &{} & -1.50 & -0.84 & -0.68 &       & 2.05  & 0.96  & 1.15  &       & 1.73  & 0.88  & 1.02  &       & 0.03  & 0.02  & 0.00 \\
          &       &{$n$ SD} &{} & 26.15 & 25.80 & 24.69 &       & 20.86 & 20.28 & 19.28 &       & 29.90 & 29.00 & 27.61 &       & 4.14  & 4.10  & 3.89 \\
          &       &{$n$ RMSE} &{} & 26.19 & 25.81 & 24.70 &       & 20.96 & 20.30 & 19.31 &       & 29.95 & 29.01 & 27.63 &       & 4.14  & 4.10  & 3.89 \\
          &       &{$n\sqrt{\bar V}$} &{} & 25.45 & 24.92 & 24.28 &       & 19.85 & 19.21 & 18.99 &       & 28.53 & 27.55 & 27.35 &       & 4.09  & 4.02  & 3.85 \\
          & $n=300$ &{$n$ Bias} &{} & -1.42 & -0.66 & -0.69 &       & 1.91  & 0.91  & 1.11  &       & 1.33  & 0.26  & 0.30  &       & -0.11 & -0.13 & -0.08 \\
          &       &{$n$ SD} &{} & 43.82 & 43.68 & 41.84 &       & 33.93 & 33.89 & 32.25 &       & 48.77 & 48.52 & 46.26 &       & 6.92  & 6.98  & 6.57 \\
          &       &{$n$ RMSE} &{} & 43.85 & 43.68 & 41.84 &       & 33.99 & 33.90 & 32.26 &       & 48.79 & 48.52 & 46.27 &       & 6.92  & 6.98  & 6.57 \\
          &       &{$n\sqrt{\bar V}$} &{} & 43.35 & 43.07 & 41.50 &       & 33.35 & 33.17 & 32.06 &       & 48.06 & 47.65 & 46.13 &       & 6.87  & 6.90  & 6.54 \\\hline
    Unstr & $n=50$ &{$n$ Bias} &{} & -1.06 & -0.46 & -0.95 &       & 1.14  & 0.19  & 1.40  &       & 1.06  & 0.09  & 0.85  &       & 0.02  & 0.02  & 0.04 \\
          &       &{$n$ SD} &{} & 21.66 & 21.13 & 21.48 &       & 25.35 & 24.47 & 25.03 &       & 35.80 & 34.61 & 35.56 &       & 7.05  & 6.76  & 6.90 \\
  $\th=1.2$        &       &{$n$ RMSE} &{} & 21.69 & 21.14 & 21.50 &       & 25.38 & 24.47 & 25.07 &       & 35.81 & 34.61 & 35.57 &       & 7.05  & 6.76  & 6.90 \\
 $\th_{12}=1.5$         &       &{$n\sqrt{\bar V}$} &{} & 20.29 & 20.27 & 20.74 &       & 23.53 & 23.48 & 24.00 &       & 33.34 & 33.32 & 34.27 &       & 6.53  & 6.52  & 6.46 \\
$\th_{13}=1.1$     & $n=100$ &{$n$ Bias} &{} & -0.72 & -0.14 & -0.49 &       & 1.63  & 0.69  & 1.98  &       & 0.70  & -0.26 & 0.18  &       & -0.04 & -0.04 & -0.07 \\
  $\th_{23}=2.7$        &       &{$n$ SD} &{} & 29.85 & 29.49 & 29.50 &       & 34.78 & 34.19 & 33.98 &       & 49.01 & 48.16 & 48.20 &       & 9.42  & 9.22  & 9.05 \\
          &       &{$n$ RMSE} &{} & 29.86 & 29.49 & 29.51 &       & 34.81 & 34.20 & 34.04 &       & 49.02 & 48.16 & 48.20 &       & 9.42  & 9.22  & 9.05 \\
          &       &{$n\sqrt{\bar V}$} &{} & 28.67 & 28.67 & 28.39 &       & 33.32 & 33.28 & 32.64 &       & 47.22 & 47.22 & 46.68 &       & 9.19  & 9.19  & 8.92 \\
     & $n=300$ &{$n$ Bias} &{} & -1.39 & -0.80 & -1.10 &       & 0.47  & -0.41 & 1.05  &       & 2.13  & 1.17  & 1.55  &       & 0.00  & 0.01  & -0.12 \\
          &       &{$n$ SD} &{} & 50.72 & 50.57 & 49.31 &       & 58.71 & 58.36 & 56.09 &       & 82.79 & 82.36 & 79.39 &       & 15.98 & 15.85 & 15.01 \\
          &       &{$n$ RMSE} &{} & 50.74 & 50.58 & 49.32 &       & 58.71 & 58.36 & 56.10 &       & 82.81 & 82.37 & 79.41 &       & 15.98 & 15.85 & 15.01 \\
          &       &{$n\sqrt{\bar V}$} &{} & 49.67 & 49.73 & 48.33 &       & 57.76 & 57.73 & 55.32 &       & 81.96 & 82.00 & 78.91 &       & 15.86 & 15.87 & 14.89 \\\bottomrule
   
    \end{tabular}
\end{footnotesize}  
  
\end{sidewaystable}

\setlength{\tabcolsep}{6pt}

\begin{table}[h]
  \centering
  \caption{\label{level}Empirical  Type I error rates for inference on $H_0:\b_4=0$ based on the weighted scores, local odds ratios GEE and ML methods. }

    \begin{tabular}{cccccccccccccc}
    \toprule
          &       &       & \multicolumn{3}{c}{$\alpha=0.01$} &       & \multicolumn{3}{c}{$\alpha=0.05$} &       & \multicolumn{3}{c}{$\alpha=0.1$} \\
    \midrule
          &       &       & WS    & GEE   & ML    &       & WS    & GEE   & ML    &       & WS    & GEE   & ML  \\\cmidrule{4-6} \cmidrule{8-10}\cmidrule{12-14}
    Exch, $\th=3$ & $n=50$ &       & 0.014 & 0.014 & 0.008 &       & 0.060 & 0.063 & 0.048 &       & 0.112 & 0.114 & 0.097 \\
          & $n=100$ &       & 0.013 & 0.013 & 0.010 &       & 0.057 & 0.056 & 0.053 &       & 0.114 & 0.112 & 0.105 \\
          & $n=300$ &       & 0.011 & 0.011 & 0.010 &       & 0.050 & 0.050 & 0.051 &       & 0.101 & 0.101 & 0.099 \\\hline
    Exch, $\th=5$ & $n=50$ &       & 0.015 & 0.019 & 0.008 &       & 0.059 & 0.068 & 0.049 &       & 0.113 & 0.120 & 0.102 \\
          & $n=100$ &       & 0.013 & 0.013 & 0.011 &       & 0.053 & 0.060 & 0.050 &       & 0.106 & 0.110 & 0.099 \\
          & $n=300$ &       & 0.012 & 0.012 & 0.010 &       & 0.054 & 0.053 & 0.051 &       & 0.105 & 0.107 & 0.102 \\\hline
    Unstr & $n=50$ &       & 0.020 & 0.015 & 0.024 &       & 0.072 & 0.060 & 0.068 &       & 0.132 & 0.117 & 0.121 \\
          & $n=100$ &       & 0.012 & 0.011 & 0.014 &       & 0.056 & 0.050 & 0.056 &       & 0.111 & 0.102 & 0.110 \\
          & $n=300$ &       & 0.010 & 0.009 & 0.009 &       & 0.050 & 0.049 & 0.051 &       & 0.101 & 0.099 & 0.105 \\
   \bottomrule
    \end{tabular}
\medskip
  
  \begin{footnotesize}
For unstructured dependence the true copula parameters are $\{\th,\th_{12},\th_{13},\th_{23}\}=\{1.2,1.5,1.1,2.7\}$.
 \end{footnotesize}
\end{table}

Table~\ref{biasmse} contains the parameter values, the bias,
standard deviations (SD) and root mean square errors (RMSE) of the
maximum likelihood (ML), weighted scores (WS) and GEE estimates,  along
with the average of their theoretical SDs ($\sqrt{\bar V}$).
The theoretical variance of the ML estimate is obtained
via the gradients and the Hessian computed
numerically during the maximization process.
The GEE estimates and their theoretical variance  are
calculated with  the function {\tt ordLORgeeR} in the R package {\tt multgee} \citep{touloumis2015}. 
{For the local odds ratios GEE approach we use the `uniform'  and the `category exchangeability' structure for the exchangeable and unstructured case, respectively, as suggested by \cite{touloumis-agresti-kateri-2013}.}
{From the results, we can see that the weighted scores and the local odds ratio  GEE method are robust to dependence and nearly as efficient as maximum likelihood for fully specified copula models.}

Furthermore,  Table \ref{level} contains  the observed level of the bilateral test for three common nominal levels  for inference on $H_0: \b_4=0$ based on the weighted scores, local odds ratios GEE and ML methods.The observed levels are close to nominal levels and hence demonstrate that the tests from all the competing approaches are reliable.

Finally in order to study the relative performance of the weighted scores over the local odds radios GEE method as the dimension $d$ or the number of categories increase we  randomly generated $B=20$ samples of size  $n = 100$ from the Gumbel copula model with exchangeable dependence  for $d,K \in \{10,15,20,25\}$. The link function, model parameters and covariates are set as before.  The simulations   were carried out on an Intel(R) Xeon(R) CPU    X5650  \@ 2.67GHz.

\setlength{\tabcolsep}{26pt}

\begin{table}[!h]
  \centering
  \caption{\label{times}Computing times (averaged over 20 replications) in seconds of the weighted scores (WS) over the   local odds ratios GEE approach.  
}
    \begin{tabular}{cccccc}
    \toprule
    $d$ & $K$ & $\theta$ & time(WS)   & time(GEE)   & $\frac{\mbox{time(GEE)}}{\mbox{time(WS)}}$  \\
    \midrule
   
    10    & 10    & 3     & 88.1  & 35.3  & 0.4 \\
          &       & 5     & 95.4  & 69.7  & 0.7 \\
    10    & 15    & 3     & 184.2 & 126.7 & 0.7 \\
          &       & 5     & 193.9 & 182.2 & 0.9 \\
    10    & 20    & 3     & 363.2 & 444.9 & 1.2 \\
          &       & 5     & 422.8 & 640.0 & 1.5 \\
    10    & 25    & 3     & 760.8 & 1559.4 & 2.0 \\
          &       & 5     & 716.1 & 1490.9 & 2.1 \\
    15    & 10    & 3     & 208.4 & 254.1 & 1.2 \\
          &       & 5     & 219.6 & 388.2 & 1.8 \\
    15    & 15    & 3     & 350.4 & 1267.6 & 3.6 \\
          &       & 5     & 371.1 & 1566.2 & 4.2 \\
    15    & 20    & 3     & 636.8 & 6057.0 & 9.5 \\
          &       & 5     & 706.1 & 6738.5 & 9.5 \\
    15    & 25    & 5     & 1524.1 & 23221.8 & 15.2 \\
          &       & 3     & 1380.5 & 25139.2 & 18.2 \\
    20    & 10    & 3     & 334.4 & 1289.0 & 3.9 \\
          &       & 5     & 360.7 & 1646.5 & 4.6 \\
    20    & 15    & 3     & 562.4 & 8165.8 & 14.5 \\
          &       & 5     & 602.9 & 9196.7 & 15.3 \\
    20    & 20    & 3     & 1287.1 & 48270.2 & 37.5 \\
          &       & 5     & 1268.0 & 50030.8 & 39.5 \\
    20    & 25    & 3     & 2131.3 & 153707.2 & 72.1 \\
          &       & 5     & 2243.1 & 138304.0 & 61.7 \\
    \bottomrule
    \end{tabular}%
  \label{tab:addlabel}%
\end{table}

Table \ref{times} summarizes the computing times (averaged over 20 replications) in seconds. Clearly the local odds ratios GEE approach requires a much higher computing time for large $d$ or $K$. Note in passing that for large $d$ or $K$ memory up to {60GB} was required for the local odds ratios GEE approach. 
Hence it is demonstrated that large matrices are  a problem for GEE methods, in as much as they cause model fitting algorithms to run slowly. {Note in passing that for larger (than the ones in Table \ref{times})  values of $K$ or $d$ the local odds ratio GEE implementation \citep{touloumis2015} is infeasible. }

\subsection{\label{sim-sel} Model selection criteria}
We perform simulation studies to examine the reliability of using CL1AIC  and CL1BIC
to choose the correct model for longitudinal ordinal data. In Subsection \ref{seccorsel} we assess the performance of CL1AIC, CL1BIC in correlation structure selection, and in  Subsection \ref{secvarsel} we investigate the performance of CL1AIC, CL1BIC in variable selection.
For exchangeable, AR(1), and unstructured  dependence,
the Gumbel copula, the mixture of max-id copula
 with Laplace transform
 $\phi_G(\cdot;\th)$ and the bivariate
Gumbel copula for the $C_{jk}^{(m)}(\cdot;\theta_{jk})$, and the MVT copula   were used as the ``true models", respectively.

\subsubsection{\label{seccorsel}Correlation structure selection}

We randomly generate $B=10^3$ samples of size  $n = 50, 100, 300$ with $d=3$  and ordinal probit regression with
$p=3, \x_{ij}=(1,x_{1ij},j-1)^T$ where $x_{1ij}$ are taken as Bernoulli random
variables with probability of success $1/2$, and $\beta_0=0.25=-\beta_1=-\beta_2$.

In Table \ref{varsel}, we present the number of times that different
working correlation structures are chosen over 1000 simulation runs under each true correlation structure.
If the true correlation structure is exchangeable or AR(1), CL1BIC  is better than CL1AIC.  
If the true correlation structure is unstructured, CL1AIC performs extremely well, especially for a small sample size, which is typical of medical studies. The difference between the correct identification rate of CL1AIC and that of CL1BIC becomes small when the sample size increases to 100 or 300.
The CL1AIC tends to choose the unstructured correlation structure more often than CL1BIC 
does, since  AIC is more likely to result in an overparametrized
model than BIC in parametric settings \citep{Chen&Lazar2012} .

\subsubsection{\label{secvarsel}Variable selection}
We randomly generate $B=10^3$ samples of size $n = 50, 100, 300$ with $d=3$   and ordinal probit regression with
$p=5, \x_{ij}=(1,x_{1ij},j-1,x_{3ij},x_{4ij})^\top$ where $x_{1ij},\beta_0,\beta_1,\beta_2$ are  as before, $x_{3ij},x_{4ij}$  are independent uniform random
variables in the interval $[-1,1]$ (and independent of $x_{1ij}$),   and $\beta_3=\beta_4=0$.
We consider the same candidate models, with various  subsets of covariates, and include all the aforementioned parametric correlation structures as  true correlation structures. The subsets of covariates that we consider are the following:
\begin{itemize}
\item $\x_{1}=(1,x_{1ij})^\top$.
\item $\x_{12}=(1,x_{1ij},j-1)^\top$ (the true regression model).
\item $\x_{123}=(1,x_{1ij},j-1,x_{3ij})^\top$.
\item $\x_{1234}=(1,x_{1ij},j-1,x_{3ij},x_{4ij})^\top$.
\end{itemize}

In Table \ref{varsel}, we present the number of times that different
subsets of covariates are chosen over 1000 simulation runs under each true correlation structure. 
For all the true correlation structures, CL1BIC performs better than CL1AIC, and its performance increases as the sample size increases.

\setlength{\tabcolsep}{7pt}
\begin{table}[!h]
  \centering
  \caption{\label{varsel}Frequencies of the correlation structure and  the set of the variables  identified identified using CL1AIC and CL1BIC from 1000 simulation runs in each setting. The first column indicates the true correlation structure;  the numbers of correct choices by each criterion are bold faced.
}
    \begin{tabular}{ccccccccccc}
    \toprule
    \multicolumn{11}{c}{Latent correlation structure selection} \\
    \hline
          &       & \multicolumn{3}{c}{$n=50$} & \multicolumn{3}{c}{$n=100$} & \multicolumn{3}{c}{$n=300$} \\
          &       & Exch  & AR(1) & Unstr & Exch  & AR(1) & Unstr & Exch  & AR(1) & Unstr \\\hline
    Exch  & CL1AIC & \textbf{732} & 123   & 145   & \textbf{779} & 63    & 158   & \textbf{814} & 3     & 183 \\
   $\th=2$       & CL1BIC & \textbf{838} & 129   & 33    & \textbf{907} & 71    & 22    & \textbf{979} & 13    & 8 \\
    AR(1) & CL1AIC & 117   & \textbf{727} & 156   & 34    & \textbf{794} & 172   & 0     & \textbf{821} & 179 \\
 & CL1BIC & 136   & \textbf{821} & 43    & 40    & \textbf{927} & 33    & 4     & \textbf{981} & 15 \\
    Unstr & CL1AIC & 64    & 13    & \textbf{923} & 64    & 13    & \textbf{923} & 0     & 0     & \textbf{1000} \\
          & CL1BIC & 249   & 34    & \textbf{717} & 51    & 4     & \textbf{945} & 0     & 0     & \textbf{1000} \\
    \hline
    \end{tabular}%
 
  \centering

    \begin{tabular}{cccccccccccccc}
    
    \multicolumn{14}{c}{Variable selection} \\
    \hline
          &       & \multicolumn{4}{c}{$n=50$}  & \multicolumn{4}{c}{$n=100$} &  \multicolumn{4}{c}{$n=300$} \\
          &       & $\x_1$ & $\x_{12}$ & $\x_{123}$ & $\x_{1234}$ & $\x_1$ & $\x_{12}$ & $\x_{123}$ & $\x_{1234}$ & $\x_1$ & $\x_{12}$ & $\x_{123}$ & $\x_{1234}$ \\\hline
    Exch  & CL1AIC & 1     & \textbf{675} & 162   & 162   & 0     & \textbf{701} & 152   & 147   & 0     & \textbf{692} & 173   & 135 \\
          & CL1BIC & 3     & \textbf{862} & 88    & 47    & 0     & \textbf{884} & 84    & 32    & 0     & \textbf{939} & 49    & 12 \\
    AR(1) & CL1AIC & 3     & \textbf{653} & 183   & 161   & 0     & \textbf{189} & 144   & 667   & 0     & \textbf{673} & 168   & 159 \\
          & CL1BIC & 22    & \textbf{819} & 107   & 52    & 2     & \textbf{876} & 87    & 35    & 0     & \textbf{924} & 55    & 21 \\
    Unstr & CL1AIC & 93    & \textbf{512} & 191   & 204   & 93    & \textbf{512} & 191   & 204   & 0     & \textbf{599} & 199   & 202 \\
          & CL1BIC & 230   & \textbf{586} & 116   & 68    & 62    & \textbf{801} & 85    & 52    & 0     & \textbf{902} & 76    & 22 \\
    \bottomrule
    \end{tabular}%
\vspace{1ex}

 \begin{footnotesize}
For exchangeable, AR(1), and unstructured dependence the true parameters are $\th=2$, $\{\th,\th_{12},\th_{13},\th_{23}\}=\{1.5,4,1,4\}$, and $\{\rho_{12},\rho_{13},\rho_{23},\nu\}=\{-0.5,-0.3,0.3,5\}$, respectively;   $\x_{1}=(1,x_{1ij})^\top$, $\x_{12}=(1,x_{1ij},j-1)^\top$ (the true regression model), 
$\x_{123}=(1,x_{1ij},j-1,x_{3ij})^\top$, and  $\x_{1234}=(1,x_{1ij},j-1,x_{3ij},x_{4ij})^\top$.
 \end{footnotesize}
\end{table}

\section{\label{sec-app}The rheumatoid arthritis data}
We illustrate the weighted scores method by re-analysing the rheumatoid arthritis data-set \citep{Bombardier-etal-1986}.  These data have  previously been used as an example for other methodological papers on GEE for ordinal regression \citep{Ware-lipsitz-1986,lipsitz-etal-94,touloumis-agresti-kateri-2013}. The data were taken from a randomized clinical trial  designed to evaluate  the effectiveness of the treatment Auranofin versus a placebo therapy for the treatment of rheumatoid arthritis. The repeated ordinal response is the self-assessment of arthritis,  classified on a five-level ordinal scale (1 = poor, . . . , 5 = very good). Patients (n=303) were randomized into one of the two treatment groups after baseline self-assessment followed during five months  of treatment with  measurements 
taken at the  first month and  every two months during treatment resulting in a maximum of 3 measurements per subject (unequal cluster sizes). The  covariates are time, baseline-assessment, age in years at baseline, sex and treatment. We treat time and baseline-assessment as categorical variables following 
\cite{touloumis-agresti-kateri-2013}. 
However, instead of testing for differences to the reference category we look at differences between adjacent categories (see, e.g., \citealp{Tutz&Gertheiss2016}). To this end we followed the coding scheme for ordinal independent variables in \cite{Walter-etal-1987}.

To select the appropriate correlation structure, we use the proposed model selection criteria in  the weighted scores estimating equations,  based on the full model with all covariates (Table \ref{selection}, correlation structure selection). Further, both logit and probit links are used for the ordinal regressions. According to CL1AIC  the correct correlation structure is the unstructured, while according to the
CL1BIC, it is exchangeable. In this example we will prefer CL1BIC since one can easily distinguish  between the various structures, as their difference in magnitude is large. This is not the case for the CL1AIC, where the differences are rather small. This was also the finding in our simulation studies, where it has been revealed that CL1AIC is more prone to select the unstructured case. Further, ordinal logistic regression is slightly better than ordinal probit regression.

Under the preferred exchangeable structure, we fit different models with different subsets of covariates, and find that the  model with time, baseline-assessment, treatment and  age,  
  has the smallest CL1AIC and CL1BIC. 
  Note  that in     
  \cite{touloumis-agresti-kateri-2013} age (and sex) have been not considered at all.  
  
\setlength{\tabcolsep}{22pt}

\begin{table}[!h]
  \centering
  \caption{\label{selection}The values of the different criteria for correlation structure selection at the full model and variable selection for the exchangeable structure for the arthritis data. The smallest value of each criterion is boldfaced.}
    \begin{tabular}{ccccc}
    \toprule
    Link  & \multicolumn{2}{c}{Probit } & \multicolumn{2}{c}{Logit} \\
    \hline
    \multicolumn{5}{l}{Correlation structure selection} \\\hline
          & CL1AIC & CL1BIC & {CL1AIC} & {CL1BIC} \\
    Exchangeable & 4280.92 & \textbf{4357.81} & 4275.09 & \textbf{4351.41} \\
    AR(1) & 4298.97 & 4374.26 & 4292.42 & 4367.20 \\
    Unstructured & \textbf{4279.97} & 4362.37 & \textbf{4273.87} & 4355.72 \\\hline
    \multicolumn{5}{l}{Variable selection} \\\hline
    time trt  baseline age sex  & 4280.92 & 4357.81 & 4275.09 & 4351.41 \\
    time trt baseline age  & \textbf{4277.91} & 4348.03 & \textbf{4273.14} & \textbf{4342.77} \\
    time trt baseline sex  & 4287.24 & 4357.54 & 4282.32 & 4352.04 \\
    time trt baseline & 4284.22 & \textbf{4347.71} & 4279.78 & 4342.81 \\
    trt baseline & 4305.09 & 4363.89 & 4298.98 & 4357.27 \\
    time trt & 4491.15 & 4529.31 & 4497.76 & 4535.93 \\
    time  & 4515.26 & 4546.00 & 4517.43 & 4548.24 \\
    trt   & 4511.37 & 4545.76 & 4517.11 & 4551.46 \\
    \bottomrule
    \end{tabular}%
  
\end{table}

Finally, Table \ref{arthritis} gives the estimates  and standard errors  of the
model parameters obtained using the  weighted scores estimating equations and GEE under the optimal exchangeable correlation structure, set of covariates, and logit link. Clearly a ``true'' copula model cannot be known for this (or any other) example, hence copula  models (e.g., Gumbel, mixture of max-id, elliptical) are not  assumed and used for ML estimation. This is precisely a reason why our method is superior compared with the ML method. Our estimating equations based on a ``working" MVN copula-based model are robust, while on the other hand ML estimates could be biased if the univariate model is correct but dependence is modelled incorrectly. The goal of this paper is not to compare copula models for a best fit, as that type of research has already been done elsewhere; see e.g., \cite{Nikoloulopoulos&karlis07BIN}.

\setlength{\tabcolsep}{7pt}

\begin{table}[!h]
  \centering
  \caption{\label{arthritis} Weighted scores and GEE estimates (Est.), along with their standard errors (SE) under the optimal correlation structure  and set  of covariates for the arthritis data.}
  
    \begin{tabular}{cccccccccc}
    \toprule
          & \multicolumn{4}{c}{Weighted scores} & {} & \multicolumn{4}{c}{GEE} \\\cmidrule{2-5} \cmidrule{7-10}
  
          & {Est. } & {se} & {$Z$} & {$p$-value} & {} & {Est. } & {se} & {$Z$} & {$p$-value} \\ \hline
    $\alpha_1$ & -2.050 & 0.638 & -3.215 & 0.001 &       & -2.081 & 0.637 & -3.268 & 0.001 \\
    $\alpha_2$ & 0.058 & 0.607 & 0.096 & 0.924 &       & 0.028 & 0.606 & 0.046 & 0.963 \\
    $\alpha_3$ & 2.021 & 0.612 & 3.305 & 0.001 &       & 1.994 & 0.610 & 3.268 & 0.001 \\
    $\alpha_4$ & 4.329 & 0.653 & 6.634 & {$<0.001$} & {} & 4.307 & 0.650 & 6.625 & {$<0.001$} \\
    $I(\mbox{time}=2,3)$ & -0.007 & 0.121 & -0.059 & 0.953 &       & 0.003 & 0.122 & 0.021 & 0.984 \\
    $I(\mbox{time}=3)$ & -0.370 & 0.113 & -3.267 & 0.001 &       & -0.365 & 0.113 & -3.220 & 0.001 \\
    trt   & -0.511 & 0.168 & -3.037 & 0.002 &       & -0.507 & 0.168 & -3.023 & 0.003 \\
    $I(\mbox{baseline}=2,3,4,5)$ & -0.620 & 0.380 & -1.631 & 0.103 &       & -0.650 & 0.380 & -1.710 & 0.087 \\
    $I(\mbox{baseline}=3,4,5)$ & -0.567 & 0.226 & -2.510 & 0.012 &       & -0.548 & 0.227 & -2.418 & 0.016 \\
    $I(\mbox{baseline}=4,5)$ & -1.369 & 0.236 & -5.790 & {$<0.001$} & {} & -1.395 & 0.236 & -5.921 & {$<0.001$} \\
    $I(\mbox{baseline}=5)$ & -1.417 & 0.403 & -3.519 & {$<0.001$} & {} & -1.389 & 0.406 & -3.424 & 0.001 \\
    age   & 0.013 & 0.008 & 1.656 & 0.098 &       & 0.014 & 0.008 & 1.736 & 0.083 \\
    \bottomrule
    \end{tabular}
  \label{tab:addlabel}
\end{table}

Our analysis shows
 that the estimates of all the parameters and their corresponding standard errors obtained from the weighted scores method are  nearly the same as those obtained from the local odds ratios GEE approach.  In fact, the columns of $p$-values for the two methods agree very closely and the same factors are found to be significant and insignificant. 
Our {study} has also revealed 
that age is of marginal  statistical significance.

This example also shows that if the correlation structure and the variables  in the mean function modelling are  correctly specified, then there is no loss in efficiency in GEE. {In fact,  if a `time exchangeability'  or a homogenous Goodman's row and column effects (`RC')  structure is assumed in the local odds ratios GEE approach \citep{touloumis-agresti-kateri-2013,touloumis2015} the age effect is statistically insignificant (results are not shown here)}; see also \cite{Nikoloulopoulos2015d} for another concrete example {for longitudinal binary  (special case of ordinal)}.  Hence, an advantage of our method  is the variable/correlation structure selection, which is well-grounded in likelihood theory, and cannot be used in GEE methods, which are based on moments with no defined likelihood.

\section{\label{sec-disc}Discussion}
In this article, we  have introduced ordinal logistic and probit regression in the weighted scores method for regression with dependent data.  Our method of combining the univariate scores for ordinal regression is theoretically sound, and gives estimates of regression parameters that are efficient and robust to dependence. {The theory  extends to any univariate regression method (such as multinomial probit), applied to dependent data, such as repeated measure multinomial (categorical non-ordinal).}

Comparing our method with  GEE for ordinal regression, we have shown  that GEE are generally efficient for inference for the regression parameters if the variable selection in the mean function modelling and the working correlation structure are correctly specified.
Composite likelihood information criteria for  both correlation structure and variable selection have been proposed to achieve this.  
However, our working MVN copula model is a proper multivariate model, and the correlations can be interpreted as latent {or polychoric \citep{Olsson-1979}} correlations; this is not the case for the GEE estimated correlation parameters, which can
also sometimes violate the Fr\'echet bounds of the feasible range of the correlation \citep{chaganty&joe06}.

{We would also like to stress that our method can allow any latent correlation structure  and is not restricted to an exchangeable or unstructured one.   For example,  SAS software only offers the independence working assumption as the only option to fit ordinal GEEs or  the dominant of the GEE methods in \cite{touloumis-agresti-kateri-2013} does not allow an  AR(1)-like association structure.} 

Last but not least, the weighted scores method  overcomes  computational  issues (matrix operations required for GEE estimation are  very slow or even infeasible for large $K$ and $d$) that occur in the  existing GEE {approaches/implementations} for ordinal longitudinal data.

\section*{Software}
R functions to implement the weighted scores method   and the CL1 information criteria for longitudinal ordinal data have been implemented in the package {\it weightedScores} \citep{nikoloulopoulos&joe11} within the open source statistical environment {\tt R} \citep{CRAN}.

\section*{Acknowledgements}
 The simulations presented in this paper were carried out on the High Performance Computing Cluster supported by the Research and Specialist Computing Support service at the University of East Anglia.

\section*{Appendix}

For $d=4$ the matrices involved in the calculation of the sensitivity matrix $\bfD_\g$ of the CL1 estimating  functions $\g$ take the form:

$$
-\bfD_{\g_1}=\X_i^\top E\begin{pmatrix}
\frac{\p\s_{i1}^{(1)}(\a)}{\p\gbf_{i1}}&0&0&0\\
0&\frac{\p\s_{i2}^{(1)}(\a)}{\p\gbf_{i2}}&0&0\\
0&0&\frac{\p\s_{i3}^{(1)}(\a)}{\p\gbf_{i3}}&0\\
0&0&0&\frac{\p\s_{i4}^{(1)}(\a)}{\p\gbf_{i4}}
\end{pmatrix}\X_i;
$$

$$
-\bfD_{\g_{2,1}}=E
\begin{pmatrix}
\frac{\p\s_{i,12}^{(2)}(\a,\rho_{12})}{\p\gbf_{i1}} & \frac{\p\s_{i,12}^{(2)}(\a,\rho_{12})}{\p\gbf_{i2}} &0 &0\\
\frac{\p\s_{i,13}^{(2)}(\a,\rho_{13})}{\p\gbf_{i1}} & 0& \frac{\p\s_{i,13}^{(2)}(\a,\rho_{13})}{\p\gbf_{i3}} &0\\
\frac{\p\s_{i,14}^{(2)}(\a,\rho_{14})}{\p\gbf_{i1}} & 0& 0&\frac{\p\s_{i,14}^{(2)}(\a,\rho_{14})}{\p\gbf_{i4}}\\
0&\frac{\p\s_{i,23}^{(2)}(\a,\rho_{23})}{\p\gbf_{i2}}&\frac{\p\s_{i,23}^{(2)}(\a,\rho_{23})}{\p\gbf_{i3}}&0\\
0&\frac{\p\s_{i,24}^{(2)}(\a,\rho_{24})}{\p\gbf_{i2}}&0&\frac{\p\s_{i,24}^{(2)}(a,\rho_{24})}{\p\gbf_{i4}}\\
0&0&\frac{\p\s_{i,34}^{(2)}(\a,\rho_{34})}{\p\gbf_{i3}}&\frac{\p\s_{i,34}^{(2)}(\a,\rho_{34})}{\p\gbf_{i4}}
\end{pmatrix}\X_i;
$$

$$
-\bfD_{\g_2}=E
\begin{pmatrix}
\frac{\p\s_{i,12}^{(2)}(\a,\rho_{12})}{\p\rho_{12}} & 0&0&0&0&0\\
0&\frac{\p\s_{i,13}^{(2)}(\a,\rho_{13})}{\p\rho_{13}} &0&0&0&0\\
0&0& \frac{\p\s_{i,23}^{(2)}(\a,\rho_{14})}{\p\rho_{14}}&0&0&0\\
0&0&0& \frac{\p\s_{i,23}^{(2)}(\a,\rho_{23})}{\p\rho_{23}}&0&0\\
0&0&0&0& \frac{\p\s_{i,24}^{(2)}(\a,\rho_{24})}{\p\rho_{24}}&0\\
0&0&0&0&0& \frac{\p\s_{i,34}^{(2)}(\a,\rho_{34})}{\p\rho_{34}}
\end{pmatrix}.
$$

The elements of these matrices are calculated as below:
$$
-E\Bigl(\frac{\p\s_{i,jk}^{(2)}(\a,\rho_{jk})}{\p\rho_{jk}}\Bigr)=
-E\Bigl(\frac{\p^2\log f_2(y_{ij},y_{ik};\nu_{ij},\nu_{ik},\gbf,\rho_{jk})}{\p\rho_{jk}^2}\Bigr)
=E\Bigl(\bigr(\frac{\p\log f_2(y_{ij},y_{ik};\nu_{ij},\nu_{ik},\gbf,\rho_{jk})}{\p\rho_{jk}}\bigl)^2\Bigr),
$$
where $\frac{\p\log f_2(y_{ij},y_{ik};\nu_{ij},\nu_{ik},\gbf,\rho_{jk})}{\p\rho_{jk}}=
\frac{\p f_2(y_{ij},y_{ik};\nu_{ij},\nu_{ik},\gbf,\rho_{jk})}{\p\rho_{jk}}/f_2(y_{ij},y_{ik};\nu_{ij},\nu_{ik},\gbf,\rho_{jk});$

\begin{eqnarray*}-E\Bigl(\frac{\p\s_{i,jk}^{(2)}(\a,\rho_{jk})}{\p\a^\top}\Bigr)
&=&-E\Bigl(\frac{\p^2\log f_2(y_{ij},y_{ik};\nu_{ij},\nu_{ik},\gbf,\rho_{jk})}{\p\a^\top\p\rho_{jk}}\Bigr)\\
&=&E\Bigl(\frac{\p\log f_2(y_{ij},y_{ik};\nu_{ij},\nu_{ik},\gbf,\rho_{jk})}{\p\a^\top}\frac{\p\log f_2(y_{ij},y_{ik};\nu_{ij},\nu_{ik},\gbf,\rho_{jk})}{\p\rho_{jk}}\Bigr),
\end{eqnarray*}
where

$
\frac{\p\log f_2(y_{ij},y_{ik};\nu_{ij},\nu_{ik},\gbf,\rho_{jk})}{\p\a^\top}=
\frac{\p f_2(y_{ij},y_{ik};\nu_{ij},\nu_{ik},\gbf,\rho_{jk})}{\p\a^\top}/f_2(y_{ij},y_{ik};\nu_{ij},\nu_{ik},\gbf,\rho_{jk})$,

$\frac{\p f_2(y_{ij},y_{ik};\nu_{ij},\nu_{ik},\gbf,\rho_{jk})}{\p\a^\top}=
\frac{\p f_{2ijk}(y_{ij},y_{ik};\gbf_{ij},\gbf_{ik},\rho_{jk})}{\p\gbf_{ij}}\X_{ij} + \frac{\p f_{2ijk}(y_{ij},y_{ik};\gbf_{ij},\gbf_{ik},\rho_{jk})}{\p\gbf_{ik}}\X_{ik}$,

$\frac{\p f_{2ijk}(y_{ij},y_{ik};\gbf_{ij},\gbf_{ik},\rho_{jk})}{\p\gbf_{ij}}=
\frac{\p f_{2ijk}(y_{ij},y_{ik};\gbf_{ij},\gbf_{ik},\rho_{jk})}{\p \Phi^{-1}\bigl(F_1(y_{ij};\gbf_{ij})\bigr)}\frac{\p\Phi^{-1}\bigl(F_1(y_{ij};\gbf_{ij})\bigr)}{\p \gbf_{ij}}+\frac{\p f_{2ijk}(y_{ij},y_{ik};\gbf_{ij},\gbf_{ik},\rho_{jk})}{\p \Phi^{-1}\bigl(F_1(y_{ij}-1;\gbf_{ij})\bigr)}\frac{\p\Phi^{-1}\bigl(F_1(y_{ij}-1;\gbf_{ij})\bigr)}{\p \gbf_{ij}}
$,

$\frac{\p\Phi^{-1}\bigl(F_1(y_{ij};\gbf_{ij})\bigr)}{\p \gbf_{ij}}=
\sum_{1}^{y_{ij}}\frac{\p f_1(y_{ij};\gbf_{ij})}{\p \gbf_{ij}}/{\phi\Bigl(\Phi^{-1}\bigl(F_1(y_{ij};\gbf_{ij})\bigr)\Bigr)}$, where 
$\frac{\p f_1(y_{ij};\gbf_{ij})}{\p \gbf_{ij}}=f_1(y_{ij};\gbf_{ij})\frac{\p\ell_{1ij}(\gbf_{ij},y_{ij})}{\p \gbf_{ij}}$.
At the above formulas
$$f_{2ijk}(y_{ij},y_{ik};\gbf_{ij},\gbf_{ik},\rho_{jk})=\int_{\Phi^{-1}[F_1(y_{ij}-1;\gbf_{ij})]}^{\Phi^{-1}[F_1(y_{ij};\gbf_{ij})]}
\int_{\Phi^{-1}[F_1(y_{ik}-1;\gbf_{ik})]}^{\Phi^{-1}[F_1(y_{ik};\gbf_{ik})]}  \phi_2(z_j,z_d;\rho_{jk}) dz_j dz_k,$$
where $F_{1ij}(y;\gbf_{ij})=\mathcal{F}(\gamma_{ijy})$.

\noindent The derivatives $\frac{\p f_2(y_{ij},y_{ik};\gbf_{ij},\gbf_{ik},\rho_{jk})}{\p\rho_{jk}}$ and $\frac{\p f_{2ijk}(y_{ij},y_{ik};\gbf_{ij},\gbf_{ik},\rho_{jk})}{\p \Phi^{-1}\bigl(F_1(y_{ij};\gbf_{ij})\bigr)}$ are computed with the  R functions {\it exchmvn.deriv.rho} and {\it exchmvn.deriv.margin}, respectively, in the  R package {\it mprobit} \citep{joe95,Joe-Wei-11}.

\baselineskip=13pt


\begin{thebibliography}{48}
\providecommand{\natexlab}[1]{#1}
\providecommand{\url}[1]{\texttt{#1}}
\expandafter\ifx\csname urlstyle\endcsname\relax
  \providecommand{\doi}[1]{doi: #1}\else
  \providecommand{\doi}{doi: \begingroup \urlstyle{rm}\Url}\fi


\itemsep=0pt
\bibitem[Achten et~al.(2010)Achten, Parsons, Edlin, Griffin, and
  Costa]{Achten-etal-2010}
{\rm Achten, J., Parsons, N., Edlin, R., Griffin, D., {\rm and} Costa, M.}
  (2010).
\newblock A randomised controlled trial of total hip arthroplasty versus
  resurfacing arthroplasty in the treatment of young patients with arthritis of
  the hip joint.
\newblock \emph{BMC Musculoskeletal Disorders}, {\bf 11}.

\bibitem[Albert and McShane(1995)]{Albert&McShane1995}
{\rm Albert, P.~S. {\rm and} McShane, L.~M.} (1995).
\newblock A generalized estimating equations approach for spatially correlated
  binary data: Applications to the analysis of neuroimaging data.
\newblock \emph{Biometrics}, {\bf 51}\penalty0 (2), \penalty0 627--638.

\bibitem[Bergsma and Rudas(2002)]{bergsma&rudas2002}
{\rm Bergsma, W.~P. {\rm and} Rudas, T.} (2002).
\newblock Marginal models for categorical data.
\newblock \emph{Ann. Statist.}, {\bf 30}, \penalty0 140--159.

\bibitem[Bombardier et~al.(1986)Bombardier, Ware, and
  Russell]{Bombardier-etal-1986}
{\rm Bombardier, C., Ware, J.~H., {\rm and} Russell, I.~J.} (1986).
\newblock Auranofin therapy and quality of life in patients with rheumatoid
  arthritis.
\newblock \emph{American Journal of Medicine}, {\bf 81}, \penalty0 565--578.

\bibitem[Chaganty and Joe({2006})]{chaganty&joe06}
{\rm Chaganty, N.~R. {\rm and} Joe, H.} ({2006}).
\newblock {Range of correlation matrices for dependent Bernoulli random
  variables}.
\newblock \emph{{Biometrika}}, {\bf {93}}\penalty0 ({1}), \penalty0 {197--206}.

\bibitem[Chen and Lazar(2012)]{Chen&Lazar2012}
{\rm Chen, J. {\rm and} Lazar, N.~A.} (2012).
\newblock Selection of working correlation structure in generalized estimating
  equations via empirical likelihood.
\newblock \emph{Journal of Computational and Graphical Statistics}, {\bf
  21}\penalty0 (1), \penalty0 18--41.

\bibitem[Crowder(1995)]{Crowder95}
{\rm Crowder, M.} (1995).
\newblock On the use of a working correlation matrix in using generalised
  linear models for repeated measures.
\newblock \emph{Biometrika}, {\bf 82}\penalty0 (2), \penalty0 407--410.

\bibitem[Dawson et~al.(1996)Dawson, Fitzpatrick, Carr, and Murray]{Dawson1996}
{\rm Dawson, J., Fitzpatrick, R., Carr, A., {\rm and} Murray, D.} (1996).
\newblock Questionnaire on the perceptions of patients about total hip
  replacement.
\newblock \emph{Journal of Bone and Joint Surgery -- Series B}, {\bf
  78}\penalty0 (2), \penalty0 185--190.

\bibitem[Heagerty and Zeger(1996)]{heagerty&zeger1996}
{\rm Heagerty, P.~J. {\rm and} Zeger, S.~L.} (1996).
\newblock Marginal regression models for clustered ordinal measurements.
\newblock \emph{Journal of the American Statistical Association}, {\bf
  91}\penalty0 (435), \penalty0 1024--1036.

\bibitem[Joe(1995)]{joe95}
{\rm Joe, H.} (1995).
\newblock Approximations to multivariate normal rectangle probabilities based
  on conditional expectations.
\newblock \emph{Journal of the American Statistical Association}, {\bf
  90}\penalty0 (431), \penalty0 957--964.

\bibitem[Joe(1997)]{joe97}
{\rm Joe, H.} (1997).
\newblock \emph{Multivariate {M}odels and {D}ependence {C}oncepts}.
\newblock Chapman \& Hall, London.

\bibitem[Joe(2014)]{joe2014}
{\rm Joe, H.} (2014).
\newblock \emph{Dependence Modeling with Copulas}.
\newblock Chapman \& Hall, London.

\bibitem[Joe and Hu(1996)]{joe&hu96}
{\rm Joe, H. {\rm and} Hu, T.} (1996).
\newblock Multivariate distributions from mixtures of max-infinitely divisible
  distributions.
\newblock \emph{Journal of Multivariate Analysis}, {\bf 57}\penalty0 (2),
  \penalty0 240--265.

\bibitem[Joe et~al.(2011)Joe, Chou, and Zhang]{Joe-Wei-11}
{\rm Joe, H., Chou, L.~W., {\rm and} Zhang, H.} (2011).
\newblock \emph{{ mprobit}: Multivariate probit model for binary/ordinal
  response}.
\newblock URL \url{http://CRAN.R-project.org/package=mprobit}.
\newblock R package version 0.9-3.

\bibitem[Johnson and Kotz(1972)]{Johnson&Kotz72}
{\rm Johnson, N.~L. {\rm and} Kotz, S.} (1972).
\newblock \emph{Continuous {M}ultivariate {D}istributions}.
\newblock Wiley, New York.

\bibitem[Larrabee et~al.(2014)Larrabee, Scott, and Bello]{Larrabee-etal-2014}
{\rm Larrabee, B., Scott, H.~M., {\rm and} Bello, N.~M.} (2014).
\newblock Ordinary least squares regression of ordered categorical data:
  Inferential implications for practice.
\newblock \emph{Journal of Agricultural, Biological, and Environmental
  Statistics}, {\bf 19}\penalty0 (3), \penalty0 373--386.

\bibitem[Lee and Nelder({2009})]{lee&nelder09}
{\rm Lee, Y. {\rm and} Nelder, J.~A.} ({2009}).
\newblock {Likelihood Inference for Models with Unobservables: Another View}.
\newblock \emph{{Stastical Science}}, {\bf {24}}\penalty0 ({3}), \penalty0
  {255--269}.

\bibitem[Liang and Zeger(1986)]{liang&zeger86}
{\rm Liang, K.~Y. {\rm and} Zeger, S.~L.} (1986).
\newblock Longitudinal data analysis using generalized linear models.
\newblock \emph{Biometrika}, {\bf 73}, \penalty0 13--22.

\bibitem[Lindsey and Lambert({1998})]{Lindsey&Lambert98}
{\rm Lindsey, J.~K. {\rm and} Lambert, P.} ({1998}).
\newblock {On the appropriateness of marginal models for repeated measurements
  in clinical trials}.
\newblock \emph{{Statistics in Medicine}}, {\bf {17}}\penalty0 ({4}), \penalty0
  {447--469}.

\bibitem[Lipsitz et~al.(1994{\natexlab{a}})Lipsitz, Kim, and
  Zhao]{lipsitz-etal-94}
{\rm Lipsitz, S.~R., Kim, K., {\rm and} Zhao, L.} (1994{\natexlab{a}}).
\newblock Analysis of repeated categorical data using generalized estimating
  equations.
\newblock \emph{Statistics in Medicine}, {\bf 13}\penalty0 (11), \penalty0
  1149--1163.

\bibitem[Lipsitz et~al.(1994{\natexlab{b}})Lipsitz, Fitzmaurice, Orav, and
  Laird]{Lipsitz-etal-1994-Biometrics}
{\rm Lipsitz, S., Fitzmaurice, G., Orav, E., {\rm and} Laird, N.}
  (1994{\natexlab{b}}).
\newblock Performance of generalized estimating equations in practical
  situations.
\newblock \emph{Biometrics}, {\bf 50}\penalty0 (1), \penalty0 270--278.

\bibitem[Nelsen(2006)]{nelsen06}
{\rm Nelsen, R.~B.} (2006).
\newblock \emph{An {I}ntroduction to {C}opulas}.
\newblock Springer-Verlag, New York.

\bibitem[Nikoloulopoulos(2013)]{Nikoloulopoulos2013a}
{\rm Nikoloulopoulos, A.~K.} (2013).
\newblock Copula-based models for multivariate discrete response data.
\newblock In {\rm Durante, F., H\"{a}rdle, W., {\rm and} Jaworski, P.},
  editors, \emph{Copulae in Mathematical and Quantitative Finance}, pages
  231--249. Springer.

\bibitem[Nikoloulopoulos(2016)]{Nikoloulopoulos2015d}
{\rm Nikoloulopoulos, A.~K.} (2016).
\newblock Correlation structure and variable selection in generalized
  estimating equations via composite likelihood information criteria.
\newblock \emph{Statistics in Medicine}, {\bf 35}, \penalty0 2377--2390.

\bibitem[Nikoloulopoulos and Joe(2015{\natexlab{a}})]{nikoloulopoulos&joe11}
{\rm Nikoloulopoulos, A.~K. {\rm and} Joe, H.} (2015{\natexlab{a}}).
\newblock \emph{{ weightedScores}: Weighted scores method for regression with
  dependent data}.
\newblock URL \url{http://CRAN.R-project.org/package=weightedScores}.
\newblock {R} package version 0.9.5.1.

\bibitem[Nikoloulopoulos and Joe(2015{\natexlab{b}})]{nikoloulopoulos&joe12}
{\rm Nikoloulopoulos, A.~K. {\rm and} Joe, H.} (2015{\natexlab{b}}).
\newblock Factor copula models for item response data.
\newblock \emph{Psychometrika}, {\bf 80}, \penalty0 126--150.

\bibitem[Nikoloulopoulos and Karlis(2008)]{Nikoloulopoulos&karlis07BIN}
{\rm Nikoloulopoulos, A.~K. {\rm and} Karlis, D.} (2008).
\newblock Multivariate logit copula model with an application to dental data.
\newblock \emph{Statistics in Medicine}, {\bf 27}, \penalty0 6393--6406.

\bibitem[Nikoloulopoulos and Karlis(2009)]{Nikoloulopoulos&karlis07FNM}
{\rm Nikoloulopoulos, A.~K. {\rm and} Karlis, D.} (2009).
\newblock Finite normal mixture copulas for multivariate discrete data
  modeling.
\newblock \emph{Journal of Statistical Planning and Inference}, {\bf 139},
  \penalty0 3878--3890.

\bibitem[Nikoloulopoulos et~al.(2011)Nikoloulopoulos, Joe, and
  Chaganty]{nikoloulopoulos&joe&chaganty10}
{\rm Nikoloulopoulos, A.~K., Joe, H., {\rm and} Chaganty, N.~R.} (2011).
\newblock Weighted scores method for regression models with dependent data.
\newblock \emph{Biostatistics}, {\bf 12}, \penalty0 653--665.

\bibitem[Nooraee et~al.(2014)Nooraee, Molenberghs, and van~den
  Heuvel]{Nooraee-etal-2014}
{\rm Nooraee, N., Molenberghs, G., {\rm and} van~den Heuvel, E.~R.} (2014).
\newblock {GEE} for longitudinal ordinal data: Comparing {R}-geepack,
  {R}-multgee, {R}-repolr, {SAS-GENMOD, SPSS-GENLIN}.
\newblock \emph{Computational Statistics \& Data Analysis}, {\bf 77}, \penalty0
  70--83.

\bibitem[Olsson(1979)]{Olsson-1979}
{\rm Olsson, F.} (1979).
\newblock Maximum likelihood estimation of the polychoric correlation
  coefficient.
\newblock \emph{Psychometrika}, {\bf 44}, \penalty0 443--460.

\bibitem[Pan(2001)]{pan2001}
{\rm Pan, W.} (2001).
\newblock Akaike's information criterion in generalized estimating equations.
\newblock \emph{Biometrics}, {\bf 57}, \penalty0 120--125.

\bibitem[Parsons et~al.(2006)Parsons, Edmondson, and
  Gilmour]{parsons-etal-2006}
{\rm Parsons, N.~R., Edmondson, R.~N., {\rm and} Gilmour, S.~G.} (2006).
\newblock A generalized estimating equation method for fitting autocorrelated
  ordinal score data with an application in horticultural research.
\newblock \emph{Journal of the Royal Statistical Society: Series C (Applied
  Statistics)}, {\bf 55}\penalty0 (4), \penalty0 507--524.

\bibitem[Parsons(2013)]{Parsons-2013-SIM}
{\rm Parsons, N.~R.} (2013).
\newblock Proportional-odds models for repeated composite and long ordinal
  outcome scales.
\newblock \emph{Statistics in Medicine}, {\bf 32}\penalty0 (18), \penalty0
  3181--3191.

\bibitem[Plackett(1965)]{plackett65}
{\rm Plackett, R.} (1965).
\newblock A class of bivariate distributions.
\newblock \emph{Journal of the American Statistical Association}, {\bf 60},
  \penalty0 516--522.

\bibitem[{R Core Team}(2015)]{CRAN}
{\rm {R Core Team}} (2015).
\newblock \emph{R: A Language and Environment for Statistical Computing}.
\newblock R Foundation for Statistical Computing, Vienna, Austria.
\newblock URL \url{https://www.R-project.org/}.

\bibitem[Sabo and Chaganty(2010)]{sabo&chaganty10}
{\rm Sabo, R. {\rm and} Chaganty, N.~R.} (2010).
\newblock What can go wrong when ignoring correlation bounds in the use of
  generalized estimating equations.
\newblock \emph{Statistics in {M}edicine}, {\bf 29}, \penalty0 2501--2507.

\bibitem[Shults et~al.(2009)Shults, Sun, Tu, Kim, Amsterdam, Hilbe, and
  Ten-Have]{Shults-etal2009}
{\rm Shults, J., Sun, W., Tu, X., Kim, H., Amsterdam, J., Hilbe, J.~M., {\rm
  and} Ten-Have, T.} (2009).
\newblock A comparison of several approaches for choosing between working
  correlation structures in generalized estimating equation analysis of
  longitudinal binary data.
\newblock \emph{Statistics in medicine}, {\bf 28}\penalty0 (18), \penalty0
  2338--2355.

\bibitem[Sklar(1959)]{sklar1959}
{\rm Sklar, M.} (1959).
\newblock Fonctions de r\'epartition \`a {$n$} dimensions et leurs marges.
\newblock \emph{Publications de l'Institut de Statistique de l'Universit\'e de
  Paris}, {\bf 8}, \penalty0 229--231.

\bibitem[Touloumis(2015)]{touloumis2015}
{\rm Touloumis, A.} (2015).
\newblock {R} package {multgee}: A generalized estimating equations solver for
  multinomial responses.
\newblock \emph{Journal of Statistical Software}, {\bf 64}\penalty0 (8),
  \penalty0 1--14.
\newblock URL \url{http://www.jstatsoft.org/v64/i08/}.

\bibitem[Touloumis et~al.(2013)Touloumis, Agresti, and
  Kateri]{touloumis-agresti-kateri-2013}
{\rm Touloumis, A., Agresti, A., {\rm and} Kateri, M.} (2013).
\newblock Gee for multinomial responses using a local odds ratios
  parameterization.
\newblock \emph{Biometrics}, {\bf 69}\penalty0 (3), \penalty0 633--640.

\bibitem[Tutz and Gertheiss(2016)]{Tutz&Gertheiss2016}
{\rm Tutz, G. {\rm and} Gertheiss, J.} (2016).
\newblock Regularized regression for categorical data.
\newblock \emph{Statistical Modelling}, {\bf 16}\penalty0 (3), \penalty0
  161--200.

\bibitem[Varin et~al.(2011)Varin, Reid, and Firth]{Varin-etal2011}
{\rm Varin, C., Reid, N., {\rm and} Firth, D.} (2011).
\newblock An overview of composite likelihood methods.
\newblock \emph{Statistica Sinica}, {\bf 21}, \penalty0 5--42.

\bibitem[Walter et~al.(1987)Walter, Feinstein, and Wells]{Walter-etal-1987}
{\rm Walter, S., Feinstein, A., {\rm and} Wells, C.} (1987).
\newblock Coding ordinal independent variables in multiple regression analyses.
\newblock \emph{American Journal of Epidemiology}, {\bf 125}, \penalty0
  319--323.

\bibitem[Wang and Carey(2003)]{wang&carey2003}
{\rm Wang, Y.-G. {\rm and} Carey, V.} (2003).
\newblock Working correlation structure misspecification, estimation and
  covariate design: Implications for generalised estimating equations
  performance.
\newblock \emph{Biometrika}, {\bf 90}\penalty0 (1), \penalty0 29--41.

\bibitem[Ware and Lipsitz(1986)]{Ware-lipsitz-1986}
{\rm Ware, J.~H. {\rm and} Lipsitz, S.~R.} (1986).
\newblock Statistical methods for the analysis of repeated categorical
  outcomes.
\newblock \emph{Proceedings of the ASA Biopharmaceutical Section}, {\bf 81},
  \penalty0 254--260.

\bibitem[Zeger and Liang(1986)]{zeger&liang86}
{\rm Zeger, S.~L. {\rm and} Liang, K.~Y.} (1986).
\newblock Longitudinal data analysis for discrete and continuous outcomes.
\newblock \emph{Biometrics}, {\bf 42}, \penalty0 121--130.

\bibitem[Zhao and Joe(2005)]{zhao&joe05}
{\rm Zhao, Y. {\rm and} Joe, H.} (2005).
\newblock Composite likelihood estimation in multivariate data analysis.
\newblock \emph{The Canadian Journal of Statistics}, {\bf 33}\penalty0 (3),
  \penalty0 335--356.

\end{thebibliography}

\end{document}